\documentclass[twocolumn,noshowpacs,noshowkeys,pra]{revtex4}%
\pdfoutput=1
\def\techreport{}
\usepackage{times}
\usepackage{amsfonts}
\usepackage{amsmath}
\usepackage{amssymb}
\usepackage{mathtools}
\usepackage{graphicx,color}%
\usepackage{subfigure}
\providecommand{\U}[1]{\protect\rule{.1in}{.1in}}
\newtheorem{theorem}{Theorem}

\newenvironment{proof}[1][Proof]{\noindent\textbf{#1.} }{\ \rule{0.5em}{0.5em}}
\DeclareMathOperator{\trace}{Tr}
\DeclareMathOperator{\Var}{Var}
\newcommand{\ket}[1]{\left| #1 \right>} %
\newcommand{\bra}[1]{\left< #1 \right|} %

\begin{document}
\title{Quantum-noise limited communication with low probability of detection}
\author{Boulat A. Bash,$^1$ Saikat Guha,$^2$ Dennis Goeckel,$^3$ and Don Towsley$^1$\thanks{This research was sponsored by the National Science Foundation under
  grants CNS-0905349, CNS-1018464 and IIS-0964094, and by the DARPA Information in a Photon 
  (InPho) program under contract number HR0011-10-C-0159.}}
\affiliation{$^1$\textit{School of Computer Science, University of Massachusetts, Amherst, Massachusetts, USA 01003},\\
$^2$\textit{Quantum Information Processing Group, Raytheon BBN Technologies, Cambridge, Massachusetts, USA 02138},\\
$^3$\textit{Electrical and Computer Engineering Department, University of Massachusetts, Amherst, Massachusetts, USA 01003}
}

\begin{abstract}
We demonstrate the achievability of a square root limit on the amount of
  information transmitted reliably and with 
  \emph{low probability of detection} (LPD) over the single-mode
  lossy bosonic channel if either the eavesdropper's measurements or 
  the channel itself is subject to the slightest amount of excess noise. 
Specifically, Alice can transmit $\mathcal{O}(\sqrt{n})$ bits to Bob over $n$ 
  channel uses such that Bob's average codeword error probability is
  upper-bounded by an arbitrarily small $\delta>0$ while a passive eavesdropper,
  Warden Willie, who is assumed to be
  able to collect all the transmitted photons that do not reach Bob, has an
  average probability of detection error that is lower-bounded by 
  $\frac{1}{2}-\epsilon$ for an arbitrarily small $\epsilon>0$.
We analyze the thermal noise and pure loss channels.
The square root law holds for the thermal noise channel even 
  if Willie employs a quantum-optimal measurement, while Bob is equipped with a 
  standard coherent detection receiver. 
We also show that LPD communication is not possible on the pure loss channel.
However, this result assumes Willie to possess an ideal receiver that is not 
  subject to excess noise.
If Willie is restricted to a practical receiver with a non-zero dark current,
  the square root law is achievable on the pure loss channel.

\end{abstract}
\maketitle

Typically wireless data transmission is secured from 
  an eavesdropping third party by a cryptographic encryption protocol.
However, there are real-life scenarios where encryption arouses suspicion and
  even theoretically robust encryption can be defeated by a determined adversary
  using a non-computational method such as side-channel analysis.
Thus, protection from interception is often
  insufficient and the adversary's ability to even \emph{detect the presence} of
  a transmission must be limited. 
This is known as \emph{low probability of detection} (LPD) communication.

While practical LPD communication on radio frequency (RF)
  channels has been explored in the context of spread-spectrum
  communications \cite[Part 5, Ch.~1]{simon94ssh}, our recent
  work~\cite{bash13squarerootjsac,bash12sqrtlawisit} addressed the fundamental 
  limits of LPD communication on an additive white Gaussian noise
  (AWGN) RF channel.
However, free-space communication at optical frequencies offers
significant advantages over RF, motivating the need to analyze
  the LPD communication capability of optical communication. 
Electromagnetic waves are quantum-mechanical %
and since modern high-sensitivity optical detection systems are limited by
noise of quantum-mechanical origin, assessing the fundamental limits of LPD
optical communication necessitates an explicit quantum analysis. 

Refs.~\cite{bash13squarerootjsac,bash12sqrtlawisit} analyze the LPD 
  communication on an AWGN channel.
This corresponds to an optical channel where: (i) transmitter 
  Alice uses ideal laser light to modulate her information, and (ii)
both the adversary Warden Willie as well as the legitimate receiver Bob use
coherent detection receivers. %
However, coherent detection receivers can be decidedly suboptimal for both the 
  intended receiver Bob and Warden Willie, and 
  thus a more general analysis of LPD communication with no structural 
  assumptions on Willie's receiver other than its realization being permissible 
  by the laws of physics is desirable. 
The sub-optimality of coherent detection is particularly
pronounced in the low photon number 
  regime~\cite{giovannetti04cappureloss, Guh11},
which is relevant to LPD communication.
It is also preferable to show the possibility of LPD communication
  when Bob is equipped with a conventional (coherent detection or 
  direct detection) optical receiver, while Willie remains quantum-powerful. 
Demonstrating  how such is possible, even on a highly lossy and noisy channel,
  is our main contribution.

In this paper we provide the fundamental scaling limits for LPD communication
  on a lossy optical channel.
We limit our analysis to pure input states since, by convexity, using mixed
  states as inputs can only deteriorate the performance (since that is
  equivalent to transmitting a randomly chosen pure state from an ensemble and 
  discarding the knowledge of that choice).
We consider two types of channels: the thermal noise and the pure loss channel.
We show that if Willie has a thermal noise channel from Alice, then meaningful
  LPD communication between Alice and Bob is possible even if
  Willie is able to collect all the transmitted photons that do not reach Bob
  and employ an arbitrarily complex receiver measurement constrained only by
  the laws of quantum physics.
On the other hand, if Willie has a pure loss channel from Alice, then
  there is a receiver he can employ that is capable of perfectly
  determining when Alice is \emph{not} transmitting.
Even though this receiver can err when Alice is transmitting, we show that
  Willie can utilize it to prevent LPD communication even when Bob is
  equipped with an optimal receiver.
However, while Willie's receiver is theoretically conceivable, it has not been 
  and is unlikely to be built.
Practical receivers suffer from dark current due to a spontaneous emission 
  process.
We thus show that LPD communication is possible 
  if Willie has a pure loss channel from Alice but is limited to 
  a direct detection receiver with non-zero dark current.

In order to state the theorems that govern the LPD scaling laws,
  we denote Willie's average error probability
  $\mathbb{P}_e^{(w)}=\frac{\mathbb{P}_{FA}+\mathbb{P}_{MD}}{2}$, where 
  $\mathbb{P}_{FA}$ is the probability
  that Willie raises a false alarm when Alice did not transmit
  and $\mathbb{P}_{MD}$ is the probability that Willie misses the detection of
  Alice's transmission. 
We say that Alice communicates to Bob \emph{reliably} when 
  Bob's average decoding error probability
  $\mathbb{P}_e^{(b)}\leq\delta$ for an arbitrary $\delta>0$ given large 
  enough $n$.
We use asymptotic notation where $f(n)=\mathcal{O}(g(n))$ denotes 
  an asymptotically tight 
  upper bound on $f(n)$, and $f(n)=o(g(n))$ and $f(n)=\omega(g(n))$ denote  
  upper and lower bounds, respectively, that are not 
  asymptotically tight \cite[Ch.~3.1]{clrs2e}.

First we present a theorem that establishes the achievability of the LPD 
  communication when Willie's capabilities are limited only by the laws of
  quantum physics but his channel from Alice is subject to thermal noise.

\begin{theorem}\emph{(Square root law for the thermal noise
channel)}\label{th:thermal}
Suppose Willie has access to an arbitrarily complex receiver measurement 
  as permissible by the laws of quantum physics
  and can capture all the photons transmitted by Alice.
Let Willie's channel from 
  Alice be subject to the noise from a thermal environment that injects 
  $N_B>0$ photons per channel use on average.
Then Alice can lower-bound $\mathbb{P}_e^{(w)}\geq\frac{1}{2}-\epsilon$ for any
  $\epsilon>0$ while reliably transmitting $\mathcal{O}(\sqrt{n})$ bits
  to Bob in $n$ channel uses even if Bob only has access to a (sub-optimal) 
  coherent detection receiver.
\end{theorem}

Next we present a partial converse to Theorem \ref{th:thermal}.
It is partial because Alice is restricted to using input states with bounded 
  photon number variance. 
However, such restriction is not onerous since this restricted set subsumes all
  physically-realizable states of a bosonic mode (such as coherent states, 
  squeezed states, number states, photon-subtracted squeezed vacuum, etc.).
We show that, under this restriction, reliable transmission of
  $\omega(\sqrt{n})$ LPD bits to Bob in $n$ channel uses is impossible.
  
\begin{theorem}\emph{(Partial converse to Theorem \ref{th:thermal})}\label{th:thermalconverse}
Suppose Alice only uses quantum states with bounded photon number variance to
  communicate with Bob.
Then, if she attempts to transmit $\omega(\sqrt{n})$ bits in $n$ channel uses, 
  as $n\rightarrow\infty$, she is either detected by Willie with 
  arbitrarily low $\mathbb{P}_e^{(w)}$ or Bob cannot decode with
  arbitrarily low error probability.
\end{theorem}

Now we show that LPD communication using \emph{any} quantum state is
  impossible when Willie has a pure loss channel from Alice and is 
  limited only by the laws of physics in his receiver measurement choice.

\begin{theorem}\emph{(No LPD communication with quantum-powerful 
  Willie on a pure loss channel)}\label{th:pureloss}
Suppose Willie has a pure loss channel from Alice and is limited only 
  by the laws of physics in his receiver measurement choice.
Then Alice cannot reliably communicate to Bob using arbitrary pure 
  states while 
  limiting $\mathbb{P}_e^{(w)}\geq\epsilon$ for any $\epsilon>0$ even if Bob 
  employs a quantum-optimal receiver.
\end{theorem}

While Theorem \ref{th:pureloss} seems to preclude Alice from using a pure
  loss channel for LPD communication, its proof requires Willie to build an 
  ideal single photon direct detection receiver that detects vacuum perfectly.
However, practical photon counting receivers are subject to ``dark clicks'',
  or photon detection events when no photons are impinging on the detector's
  active surface.
We show that in this case LPD communication is possible.

\begin{theorem}\emph{(Square root law when Willie experiences dark
current)}\label{th:dark}
Suppose that Willie has a pure loss channel from Alice but is limited to
  a receiver with a non-zero dark current.
Then Alice can lower-bound $\mathbb{P}_e^{(w)}\geq\frac{1}{2}-\epsilon$ for any
  $\epsilon>0$ while reliably transmitting $\mathcal{O}(\sqrt{n})$ bits
  to Bob in $n$ channel uses.
\end{theorem}

We start this letter by introducing our optical channel model and 
  hypothesis testing. 
We then prove Theorems \ref{th:thermal}, \ref{th:thermalconverse},
  \ref{th:pureloss}, and \ref{th:dark} in succession, and conclude the letter.
\ifx\techreport\undefined \else
\section{Prerequisites}
\fi
\textit{Channel model}---Consider a single spatial mode free space
optical channel, where each channel use corresponds to one signaling interval
that carries one modulation symbol. 
We focus on
single-mode quasi-monochromatic propagation, since our results readily
generalize to multiple spatial modes (near-field link) and/or a wideband
channel with appropriate power-allocation across spatial modes and
frequencies~\cite{Sha05}. 
For simplicity of exposition we limit our analysis to vacuum
propagation, i.e., we do not address the effect of atmospheric turbulence.
The Heisenberg-picture input-output relationship of the single-mode bosonic
channel is captured by a `beamsplitter' relationship, ${\hat b} =
\sqrt{\eta}\,{\hat a} + {\sqrt{1-\eta}}\,{\hat e}$, where ${\hat a}$ and ${\hat
b}$ are modal annihilation operators of the input and output modes
respectively, and $\eta \in [0, 1]$ is the power transmissivity, the fraction
of power Alice puts in the input mode that couples into Bob's aperture.
Classically, a power attenuation is captured by the relationship $b =
{\sqrt{\eta}}\, a$, where $a$ and $b$ are complex field amplitudes of the input
and output mode functions. The quantum description of the channel requires the
`environment' mode ${\hat e}$ in order to preserve the commutator brackets,
i.e., $\left[ {\hat b}, {\hat b}^\dagger \right]=1$, which translates to
preserving the Heisenberg uncertainty relationship of quantum mechanics. For
the pure loss channel, the environment mode $\hat e$ is in a {\em vacuum}
state, i.e., ${\hat \rho}^E = |0\rangle\langle 0 |^E$. The vacuum state
captures the minimum amount of noise that must be injected when `nothing
happens' other than pure power attenuation. For a thermal noise channel, $\hat
e$ is in a thermal state with mean photon number $N_B>0$, 
  i.e.~$\hat{\rho}^E=\hat{\sigma}^T(N_B)$ where $\hat{\sigma}^T(N_B)$ is 
  a mixture of coherent states weighted by a Gaussian distribution:
\begin{align}
{\hat \sigma}^T(\bar{n}) = \sum_{i=0}^{\infty}\frac{{\bar{n}}^i}{(1+\bar{n})^{1+i}}|i\rangle \langle i|^E = \int_{\mathbb C}\frac{e^{-\frac{|\alpha|^2}{\bar{n}}}}{\pi \bar{n}}|\alpha\rangle\langle\alpha |^E{\rm d}^2\alpha. \label{eq:TS}
\end{align}
The mean number of photons injected by the thermal environment is
  $N_B \approx \pi 10^6 \lambda^3 N_\lambda / {\hbar} \omega^2$, where 
  $N_\lambda$ is the background spectral radiance 
  (in W/m$^2$ sr-$\mu$m)~\cite{Kop70}. 
A typical daytime value $N_\lambda \approx 10$ W/m$^2$ sr-$\mu$m at 
  $\lambda = 1.55 \mu$m leads to $N_B \approx 10^{-6}$ photons/mode. 
For $N_B = 0$, the thermal noise channel reduces to the pure loss channel. 

\textit{Hypothesis Testing}---Willie collects part of the transmitted light 
  during the transmission of Alice's $n$ modulation symbols and performs 
  a hypothesis test on whether Alice transmitted or not.
Willie's null hypothesis $H_0$ is that
  Alice does not transmit, and thus he observes vacuum plus noise photons,
  injected either by a thermal environment or due to dark current generated by
  a spontaneous emission process in his own measurement apparatus.
His alternate hypothesis $H_1$ is that Alice transmits.

\ifx\techreport\undefined
\textit{Thermal Noise Channel ($N_B>0$)}---We 
\else
\section{Thermal Noise Channel ($N_B>0$)}
We 
\fi
  begin by providing a constructive proof of achievability of 
  $\mathcal{O}(\sqrt{n})$ LPD
  bits in $n$ channel uses: we describe Alice and Bob's communication 
  system and prove that Willie's average probability of detection error is
  lower-bounded arbitrarily close to $\frac{1}{2}$, while Bob's average 
  probability 
  of codeword decoding error is upper-bounded arbitrarily close to zero.

\begin{proof}\emph{(Theorem \ref{th:thermal}).}
\textit{Construction:} 
Let Alice use a zero-mean isotropic Gaussian-distributed coherent state input
  $\left\{p(\alpha),\ket{\alpha}\right\}$, where $\alpha \in {\mathbb C}$, 
  $p(\alpha) = e^{-|\alpha|^2/{\bar n}}/{\pi {\bar n}}$ 
  with mean photon number per symbol
  $\bar{n}=\int_{\mathbb C}|\alpha|^2 p(\alpha){\rm d}^2\alpha$. 
Alice encodes $M$-bit blocks of input into codewords of length
  $n$ symbols at the rate $R=M/n$ bits/symbol by generating $2^{nR}$ codewords 
  $\{\bigotimes_{i=1}^n\ket{\alpha_i}_k\}_{k=1}^{2^{nR}}$, each
  according to $p(\bigotimes_{i=1}^n\ket{\alpha_i})=\prod_{i=1}^{n}p(\alpha_i)$,
  where $\bigotimes_{i=1}^n\ket{\alpha_i}=\ket{\alpha_1\ldots\alpha_{n}}$ is an
  $n$-mode tensor-product coherent state.
The codebook is used only once to send a single message and is kept secret from 
  Willie, though he knows how it is constructed.\footnote{Conceptually, the codebook is similar to a one-time pad \cite{shannon49sec}
  and the shared secret requirement follows `best practices' in security
  system design where the security of the system depends only on the secret key
  \cite{menezes96HAC}.}

\textit{Analysis (Willie):}
Suppose that Willie 
  captures all of Alice's transmitted energy that does not reach Bob's receiver.
This is a fairly strong assumption for a line-of-sight diffraction-limited 
  far-field optical link.
Since Willie does not have access to Alice's codebook,
  the $n$-channel use average quantum states at Willie's receiver under the two hypotheses are given respectively by the density operators,
\begin{align}
\hat{\rho}_0^{\otimes n}&=\left(\sum_{i=0}^\infty \frac{(\eta N_B)^i}{(1+\eta N_B)^{1+i}}\ket{i}\bra{i}\right)^{\otimes n}, \;{\text{and}}\label{eq:rho0}\\
\hat{\rho}_1^{\otimes n}&=\left(\sum_{i=0}^\infty \frac{((1-\eta)\bar{n}+\eta N_B)^i}{(1+(1-\eta)\bar{n}+\eta N_B)^{1+i}}\ket{i}\bra{i}\right)^{\otimes n}.\label{eq:rho1}
\end{align}
The quantum-limited minimum average probability of error in discriminating the 
  $n$-copy states $\hat{\rho}_0^{\otimes n}$ and $\hat{\rho}_1^{\otimes n}$ is:
\begin{align}
\label{eq:P_e}\mathbb{P}_{e,\min}^{(w)}&= \frac12\left[1 - \frac12\| \hat{\rho}_1^{\otimes n} - \hat{\rho}_0^{\otimes n} \|_1\right],
\end{align}
where $\|\hat{\rho}-\hat{\sigma}\|_1$ is the \emph{trace distance} between states
  $\hat{\rho}$ and $\hat{\sigma}$.
We can lower-bound\footnote{Since $\hat{\rho}_0$ and $\hat{\rho}_1$ are 
  diagonal in the number basis, Willie's quantum-optimal measurement to
  discriminate $\hat{\rho}_0^{\otimes n}$ and $\hat{\rho}_1^{\otimes n}$ is an ideal
  photon number resolving direct detection receiver
  with POVM elements given by the photon number operators
  $\left\{|i\rangle \langle i|\right\}$, $i \in \left\{0, 1, \ldots \right\}$. 
  We can derive $\mathbb{P}_{e, \min}^{(w)}$ exactly, however,
  Pinsker's Inequality is simple and sufficient for the bound 
  we need.}
  $\mathbb{P}_{e,\min}^{(w)}$ using quantum Pinsker's 
  Inequality \cite[Th.~11.9.2]{wilde11quantumit}:
\begin{align}
\|\hat{\rho}-\hat{\sigma}\|_1&\leq\sqrt{2D(\hat{\rho}\|\hat{\sigma})},
\end{align}
where $D(\hat{\rho}\|\hat{\sigma})\equiv\trace\{\hat{\rho}(\ln(\hat{\rho})-\log(\hat{\sigma}))\}$ 
  is the quantum relative entropy (QRE) between states $\hat{\rho}$ and $\hat{\sigma}$.
We thus have:
\begin{align}
\mathbb{P}_{e}^{(w)}&\geq \mathbb{P}_{e,\min}^{(w)}\geq\frac{1}{2}-\sqrt{\frac{1}{8}D(\hat{\rho}_0^{\otimes n}\|\hat{\rho}_1^{\otimes n})}.
\end{align}
Since QRE is additive for tensor product states, 
$D(\hat{\rho}_0^{\otimes n}\|\hat{\rho}_1^{\otimes n})=nD(\hat{\rho}_0\|\hat{\rho}_1)$. Since $\hat{\rho}_0$ and $\hat{\rho}_1$ are diagonal in the photon-number basis, the QRE is:
\begin{align}
\nonumber D(\hat{\rho}_0\|\hat{\rho}_1)&=\eta N_B\ln\frac{(1+(1-\eta)\bar{n}+\eta N_B)\eta N_B}{((1-\eta)\bar{n}+\eta N_B)(1+\eta N_B)}+\\
\label{eq:kl_therm}&\phantom{=}\;+\ln\frac{1+(1-\eta)\bar{n}+\eta N_B}{1+\eta N_B}.
\end{align}
The details of the derivation of \eqref{eq:kl_therm} are given in
  the Supplement.
The first two terms of the Taylor series expansion of \eqref{eq:kl_therm}
  around $\bar{n}=0$ are zero and the fourth term is negative.
Thus, using Taylor's Theorem we can upper-bound \eqref{eq:kl_therm} by the third term as follows:
\begin{align}
D(\hat{\rho}_0\|\hat{\rho}_1)&\leq\frac{(1-\eta)^2\bar{n}^2}{2\eta N_B (1+\eta N_B)}.
\end{align}
Therefore, setting 
\begin{align}
\label{eq:nbar}\bar{n}&=\frac{4\epsilon\sqrt{\eta N_B(1+\eta N_B)}}{\sqrt{n}(1-\eta)}
\end{align}
  ensures that Willie's error probability is lower-bounded by
  $\mathbb{P}_{e}^{(w)}\geq\frac{1}{2}-\epsilon$ over $n$ optical channel uses
  by Alice.

\textit{Analysis (Bob):} 
Suppose Bob uses a coherent detection receiver. A homodyne receiver, which is more efficient than a heterodyne receiver in the low photon number regime~\cite{giovannetti04cappureloss}, induces an AWGN channel with noise power 
  $\sigma_{b}^2=\frac{2(1-\eta)N_B+1}{4\eta}$.
Since Alice uses Gaussian modulation with symbol power $\bar{n}$ defined
  in \eqref{eq:nbar}, we can upper-bound $\mathbb{P}_e^{(b)}$ as follows
  \cite[Eq.~(7)]{bash13squarerootjsac}:
\begin{align}
\label{eq:P_e_hom}\mathbb{P}_{e}^{(b)}&\leq\delta=2^{B_{\text{hom}}(n,\epsilon,\delta)-\frac{n}{2}\log_2\left(1+\bar{n}/2\sigma_{b,\text{hom}}^2\right)}.
\end{align}
Substituting the expression for $\bar{n}$ from \eqref{eq:nbar} and 
  $\sigma_{b}^2$, and solving for the maximum number of bits 
  $B_{\text{hom}}(n,\epsilon,\delta)$ that
  can be transmitted from Alice to Bob in $n$ channel uses, we obtain:
\begin{align}
\label{eq:B_hom}B_{\text{hom}}(n,\epsilon,\delta)&={C}_d(\delta)+
\sqrt{n}{C}_c(\epsilon, \eta, N_B)+\mathcal{O}(1),
\end{align}
where ${C}_d(\delta)=\log_2\delta$ is the `cost' of upper-bounding
  Bob's decoding error probability by $\mathbb{P}_{e}^{(b)}\leq\delta$, and
  ${C}_c(\epsilon, \eta, N_B)=\frac{\epsilon\sqrt{\eta N_B(1+\eta
  N_B)}}{(1-\eta)}\times\frac{4\eta}{2(1-\eta)N_B+1}$
  is the cost of lower-bounding Willie's probability of detection by
  $\mathbb{P}_{e,\min}^{(w)}\geq \frac{1}{2}-\epsilon$.
\end{proof}

\noindent \emph{Remark.} Eq.~\eqref{eq:B_hom} illustrates that while the cost 
  of reducing Bob's decoding
  error has an additive impact that is insignificant
  at large enough $n$, the cost of limiting Willie's detection
  capabilities is multiplicative and proportional to $\epsilon$. 
\ifx\techreport\undefined \else
We plot $B_{\text{hom}}(n,\epsilon,\delta)$ using transmissivity $\eta=0.1$
  for various values of $\epsilon$
  and $\delta$ on Figure \ref{fig:B_hom}, illustrating the square root 
  law and that the probability of decoding error imposed on Bob 
  has insignificant impact, while the tolerance of being detected by Willie
  greatly affects the amount of information that can be covertly transmitted.
The small number of bits that can be sent across the channel (200 bits
  in 10,000 seconds, or roughly 2 hours 45 minutes, with $\epsilon=0.1$)
  is likely due to the very conservative assumptions we make 
  on Willie's capability.

\begin{figure*}[t]
\centering
\subfigure[~$\delta=0.01$]{\label{fig:hom_d01}\includegraphics[scale=0.60]{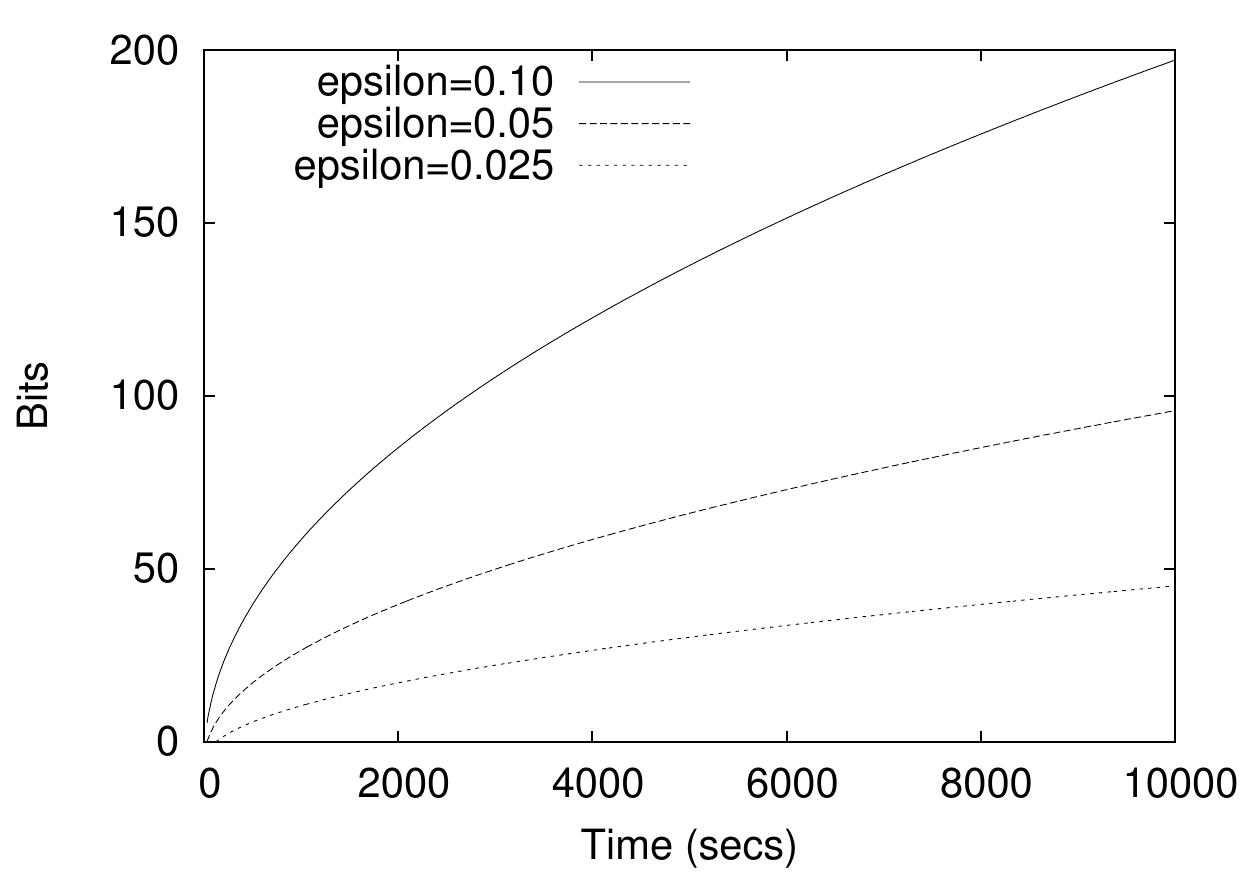}}\qquad
\hspace{10mm}
\subfigure[~$\delta=0.10$]{\label{fig:hom_d10}\includegraphics[scale=0.60]{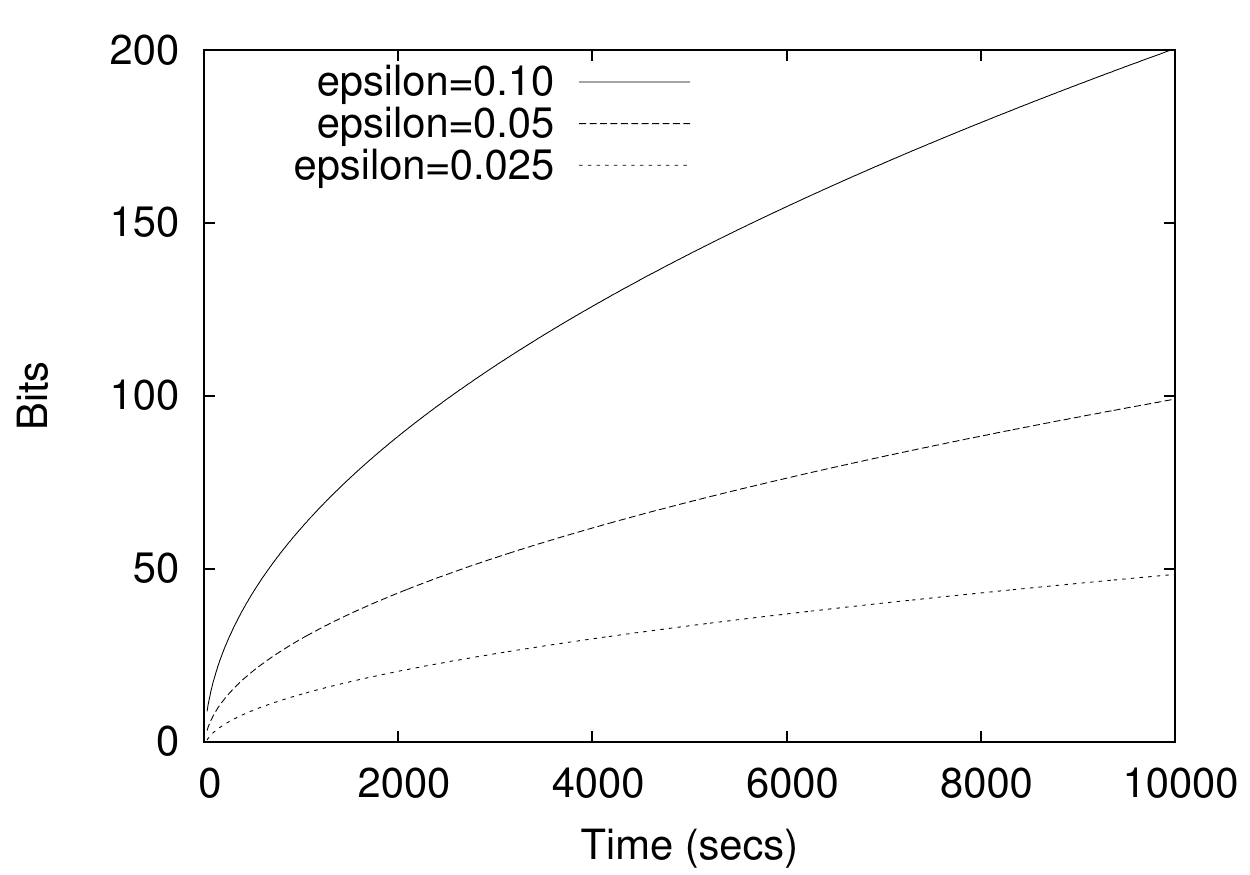}}
\caption{\label{fig:B_hom}$B_{\text{hom}}(n,\epsilon,\delta)$
  plotted for $\eta=0.1$ and several values of $\epsilon$ and $\delta$.
Here $N_B=10^{-6}$ and each modulation symbol duration is $100 \text{ps}$.  
Figures clearly illustrate that while the choice of $\delta$ is hardly 
  noticeable, choice of $\epsilon$ has a significant multiplicative impact on 
  the number of covert bits that can be sent across the channel.}
\end{figure*}
\fi

\ifx\techreport\undefined
\textit{Partial Converse to Theorem \ref{th:thermal}}---Here
\else
\section{Partial Converse to Theorem \ref{th:thermal}}
Here
\fi
Alice's objective is to transmit a message $W_k$ that is
  $M=\omega(\sqrt{n})$ bits long to Bob at the rate $R=M/n$ bits/channel use
  using a codeword containing $n$ pure
  states with arbitrarily small probability of decoding
  error as $n$ gets large, while limiting Willie's ability to detect
  her transmission.
For an upper bound on the reduction in entropy, the messages are
  chosen equiprobably.
We now show that
  if Alice violates the square root law by attempting to transmit 
  $\omega(\sqrt{n})$ bits in $n$ channel uses, as $n\rightarrow\infty$,
  she is either detected by Willie with arbitrarily low $\mathbb{P}_e^{(w)}$
  or Bob's probability of decoding error is lower-bounded by
  a positive constant.
We restrict Alice to transmitting only the states with bounded photon number
  variance, that is, for any state  $\ket{\psi}=\sum_{k=0}^\infty b_k\ket{k}$
  that Alice uses, we require that 
  $\sum_{k=0}^\infty k^2|b_k|^2\leq \sigma^2_{UB}<\infty$.
While we note that all practical states meet this requirement, in the future 
  we would like generalize this result to arbitrary pure states.

Willie uses a simple heterodyne receiver to detect Alice's transmissions.
We demonstrate that this is enough to detect with arbitrarily small 
  error probability as $n\rightarrow\infty$ those codewords with mean photon
  number per symbol $\bar{n}=\omega(1/\sqrt{n})$.
We then use Fano's inequality to show that when Alice attempts to transmit
  $\omega(\sqrt{n})$ bits 
  of information, while preventing the upper bound on the error probability
  of Willie's heterodyne receiver from being arbitrarily close to zero,
  Bob suffers non-zero decoding error probability.

\begin{proof}\emph{(Theorem \ref{th:thermalconverse}).}
Suppose Alice uses a codebook $\{\Omega_u^A,u=1,\ldots,2^{nR}\}$, where a state 
  $\Omega_u^A=\bigotimes_{i=1}^n\hat{\rho}_i^A(u)$ encodes message $W_u$ out 
  of $M$ possible messages, with
  $\hat{\rho}^A_i(u)=\ket{\psi_i(u)}\bra{\psi_i(u)}$ and
  $\ket{\psi_i(u)}=\sum_{k=0}^{\nu_i(u)}b^{(i)}_k(u)\ket{k}$ where 
  $\nu_i(u)$ can, in principle, be infinite.
First we analyze Willie's detector and assume that an arbitrary message $W_a$ 
  was transmitted.
At each channel use, Willie observes an output state $\hat{\rho}^W_i(a)$ of a thermal
  noise channel from Alice, where the channel is described by a beamsplitter
  relationship $\hat{w}=\sqrt{\gamma}\hat{a}+\sqrt{1-\gamma}\hat{e}$
  with $\hat{a}$ and $\hat{e}$ being the input and environment modes and
  $0<\gamma\leq 1-\eta$.
We subsume any sub-unity detection efficiency of Willie's heterodyne receiver
  in $\gamma$.
Then Willie's hypothesis test reduces to choosing between the states,
\begin{align}
\hat{\rho}_0^{\otimes n}&=\left(\sum_{i=0}^\infty \frac{((1-\gamma) N_B)^i}{(1+(1-\gamma) N_B)^{1+i}}\ket{i}\bra{i}\right)^{\otimes n}, \;{\text{and}}\\
\hat{\rho}_1^{\otimes n}&=\bigotimes_{i=1}^n\hat{\rho}^W_i(a)
\end{align}
where $\hat{\rho}^W_i(a)$ is the output state of a thermal noise channel with 
  transmissivity $\gamma$ corresponding to an input state $\hat{\rho}^A_i(a)$.

Willie uses a heterodyne receiver and only considers
  the squared magnitude of the complex output of this receiver 
  (thus discarding the in-phase component of his readings).
After collecting a sequence of $n$ such observations of his channel from Alice
  $\{|y_1|^2,\ldots,|y_n|^2\}$, Willie compares their average
  $S=\frac{1}{n}\sum_{i=1}^n|y_i|^2$ to a threshold.
The probability distribution for the test statistic $S$ depends on which
  hypothesis is true: we denote by $\mathbb{P}_0$ the distribution when 
  $H_0$ holds with Alice not transmitting, and $\mathbb{P}^{(a)}_1$ when
  $H_1$ holds with Alice transmitting message $W_a$.
We first show that Willie's error probabilities $\mathbb{P}_{FA}$ and
  $\mathbb{P}_{MD}$ can be bounded for this receiver given Alice's 
  codeword parameters.
Then we show that if Alice uses a codebook that makes this bound fail,
  Bob cannot decode her transmissions without error even with 
  an quantum-optimal receiver.

The statistics of heterodyne receiver measurements are given by the 
  Husimi Q 
  representation $Q(\alpha)=\frac{1}{\pi}\bra{\alpha}\hat{\rho}\ket{\alpha}$
  of the received quantum state $\hat{\rho}$.
If the null hypothesis is true and Alice is not transmitting, then Willie
  observes a sequence of attenuated thermal states, each with mean photon
  number $(1-\gamma)N_B$.
Each squared magnitude of the heterodyne receiver reading is independently
  and identically distributed (i.i.d.) and
  the Q-function of the attenuated thermal state is
  $Q^T(\alpha)=\frac{1}{\pi(1+N_B)}e^{-|\alpha|^2/(1+N_B)}$.
Therefore, under the null hypothesis, $\mathbb{E}[S]=1+(1-\gamma)N_B$ and
  $\Var[S]=\frac{(1+(1-\gamma)N_B)^2}{n}$.
Since the test statistic $S$ should be close to $1+(1-\gamma)N_B$ when Alice 
  is not transmitting, Willie picks a threshold $t$ and compares $S$ to 
  $1+(1-\gamma)N_B+t$.
Using the Chebyshev's inequality, we can upper bound the probability of the
  false alarm as follows:
\begin{align}
\mathbb{P}_{FA}&=\mathbb{P}_0(S\geq 1+(1-\gamma)N_B+t)\\
&\leq \mathbb{P}_0(|S-(1+(1-\gamma)N_B)|\geq t)\\
&\leq \frac{(1+(1-\gamma)N_B)^2}{nt^2}
\end{align}
Thus, to obtain desired $\mathbb{P}_{FA}^*$, Willie sets $t=\frac{d}{\sqrt{n}}$,
  where $d=\frac{1+(1-\gamma)N_B}{\sqrt{\mathbb{P}_{FA}^*}}$.
Note that the threshold decreases with more observations.

Now, when Alice transmits a codeword 
  $\Omega_a^A=\bigotimes_{i=1}^n\hat{\rho}_i^A(a)$,
  Willie receives the output state $\bigotimes_{i=1}^n\hat{\rho}^W_i(a)$ of the 
  thermal noise channel with transmissivity $\gamma$.
Since the output state is a tensor product, the heterodyne detector readings
  are independent but not identical.
The expected squared magnitude of each reading is:
\begin{align}
\mathbb{E}[|y_i|^2]&=\int_{\mathbb{C}}|\alpha|^2Q^W_{\ket{\psi_i(a)}}(\alpha)d^2\alpha\\
\label{eq:exp_het_read_h1}&=1+(1-\gamma)N_B+\gamma\bar{n}_i(a)
\end{align}
where $\bar{n}_i(a)=\sum_{k=0}^{\nu_i(a)}kb_k^{(i)}(a)$ denotes the mean 
  photon number of state $\hat{\rho}_i^A(a)$ and 
  $Q^W_{\ket{\psi_i(a)}}(\alpha)$ is the Q representation of 
  $\hat{\rho}^W_i(a)$.
Similarly, the variance is:
\begin{align}
\label{eq:var_het_read_h1}\Var[|y_i|^2]&=\gamma^2\sigma^2_i(a)+c_1\bar{n}_i(a)+c_2
\end{align}
where $\sigma^2_i(a)=\mu^{(2)}_i(a)-(\bar{n}_i(a))^2$ denotes the 
  photon number variance of $\hat{\rho}_i^A(a)$, and
  $c_1=2\gamma((2+N_B)(1-\gamma)-1)$, $c_2=(1+(1-\gamma)N_B)^2$.
To obtain $Q^W_{\ket{\psi_i(a)}}(\alpha)$, we convolve \footnote{Since
  the Q representation is not a probability distribution but a quasiprobability,
  the standard convolution law for the probability distributions does not 
  apply.
Given the beamsplitter relationship 
  $\hat{w}=\sqrt{\gamma}\hat{a}+\sqrt{1-\gamma}\hat{e}$ between the input 
  modes $a$ and $b$, and the output mode $w$, the Husimi Q function
  $Q_w(\alpha)=\frac{1}{1-\gamma}\int_{\mathbb{C}}Q_a(\beta)Q_b\left(\frac{\alpha-\sqrt{\gamma}\beta}{\sqrt{1-\gamma}}\right)d^2\beta$
  \cite[Eq.~(2.17)]{kim95beamsplitter}.}
  the Q 
  representation of the thermal environment $Q^T(\alpha)$ with that of 
  the input state $\ket{\psi_i(a)}$, 
  \begin{align}
Q^A_{\ket{\psi_i(a)}}(\alpha)&=\frac{1}{\pi}\sum_{k=0}^{\nu_i(a)}\sum_{l=0}^{\nu_i(a)}b^{(i)}_k(a)\left(b^{(i)}_l(a)\right)^*\frac{(\alpha^*)^k\alpha^l}{\sqrt{k!l!}}e^{-|\alpha|^2},
\end{align}
  using \cite[Eq.~(2.17)]{kim95beamsplitter}, with the details of the 
  derivation of \eqref{eq:exp_het_read_h1} and \eqref{eq:var_het_read_h1}
  in the supplement.
Since the photon number variance of $\hat{\rho}_i^A(a)$ is bounded by $\sigma^2_{UB}$,
  we have $\sigma^2_i(a)\leq \sigma^2_{UB}$.
Denoting the average photon number of the codeword $\Omega_a$ by 
  $\bar{n}(a)=\frac{1}{n}\sum_{i=1}^n\bar{n}_i(a)$, the probability of 
  missing the detection of codeword $\Omega_a$ can thus be bounded 
  using Chebyshev's inequality as follows:
\begin{align}
\mathbb{P}_{MD}^{(a)}&=\mathbb{P}^{(a)}_1(S<1+(1-\gamma)N_B+t)\\
&\leq \mathbb{P}^{(a)}_1(|S-1-(1-\gamma)N_B-\gamma\bar{n}(a)|\geq\gamma\bar{n}(a)-t)\\
&\leq \frac{\sum_{i=1}^n \gamma^2\sigma^2_i(a)+c_1\bar{n}_i(a)+c_2}{n^2\left(\gamma\bar{n}(a)-t\right)^2}\\
\label{eq:p_md_ub}&\leq\frac{\gamma\sigma^2_{UB}+c_1\bar{n}(a)}{\left(\gamma\sqrt{n}\bar{n}(a)-d\right)^2}
\end{align}
If the average photon number $\bar{n}(a)=\omega(1/\sqrt{n})$, 
  $\lim_{n\rightarrow\infty}\mathbb{P}_{MD}^{(a)}=0$.
Thus, given enough observations, Willie can detect Alice's codewords that have
  the average photon number $\bar{n}(a)=\omega(1/\sqrt{n})$ with
  arbitrarily low probability of error $\mathbb{P}_e^{(w)}$.
Note that not only is Willie oblivious to any details about Alice's codebook
  construction, but he also only needs a simple heterodyne detector to detect
  Alice.

Now, only when the transmitted codeword has average photon number 
  $\bar{n}_{\mathcal{U}}=\mathcal{O}(1/\sqrt{n})$, the upper bound in 
  \eqref{eq:p_md_ub} fails to approach zero as $n\rightarrow\infty$.
In other words, if Alice wants to lower-bound $\mathbb{P}_e^{(w)}$, her codebook
  must contain a positive fraction $\kappa$ of such low photon number codewords.
Denote the subset of messages that have codewords with the average photon number
  $\bar{n}_{\mathcal{U}}=\mathcal{O}(1/\sqrt{n})$ by $\mathcal{U}$. 
Let's examine Bob's probability of decoding error $\mathbb{P}_e^{(b)}$.
Denote by $E_{a\rightarrow k}$ the event that a transmitted message $W_a$ is 
  decoded as $W_k\neq W_a$.
Since the messages are 
  equiprobable, the average probability of error for the 
  codebook containing only the codewords in $\mathcal{U}$ is as follows: 
\begin{align}
\mathbb{P}_e^{(b)}(\mathcal{U})&=\frac{1}{|\mathcal{U}|}\sum_{W_a\in\mathcal{U}}\mathbb{P}\left(\cup_{W_k\in\mathcal{U}\backslash\{W_a\}}E_{a\rightarrow k}\right),
\end{align}
where $|\cdot|$ is the set cardinality operator.
The probability of Bob's decoding error is lower-bounded by 
  $\mathbb{P}_e^{(b)}\geq\kappa\mathbb{P}_e^{(b)}(\mathcal{U})$,
  since the equality holds only when Bob errorlessly receives messages that are
  not in $\mathcal{U}$ and knows when the messages from 
  $\mathcal{U}$ are sent (in other words, the equality holds with the set of 
  messages on which decoder can err is reduced to $\mathcal{U}$).
The probability that a message is sent from $\mathcal{U}$ is $\kappa$,
  which means that if Alice's coding rate is $R$, then there are 
  $\kappa 2^{nR}$ messages in $\mathcal{U}$.
Denote by $W_a\in\mathcal{U}$ the message transmitted by Alice, and by 
  $\hat{W}_a$ Bob's decoding of $W_a$.
Then, since each message is equiprobable,
\begin{align}
\log_2\kappa+nR&=H(W_a)\\
\label{eq:altproof_2}&=I(W_a;\hat{W}_a)+H(W_a|\hat{W}_a)\\
\label{eq:altproof_fano}&\leq I(W_a;\hat{W}_a)+1+(\log_2 \kappa+nR)\mathbb{P}_e^{(b)}(\mathcal{U})\\
\label{eq:altproof_holevo}&\leq\chi(\{\frac{1}{|\mathcal{U}|};\Omega^A_u\})+1+(\log_2\kappa+ nR)\mathbb{P}_e^{(b)}(\mathcal{U})
\end{align}
where \eqref{eq:altproof_2} is from the definition of mutual information,
  \eqref{eq:altproof_fano} is due to classical Fano's inequality
  \cite[Eq.~(9.37)]{cover02IT},
  and \eqref{eq:altproof_holevo} is the Holevo's bound 
  $I(X;Y)\leq \chi(\{p_i,\hat{\rho}_i\})$, with $\chi(\{p_i,\hat{\rho}_i\})$
  being the Holevo information for a channel with input alphabet $X$,
  $\{p_i,\hat{\rho}_i\}$ the priors and the modulating states, and $Y$ the
  resulting output alphabet (assuming a POVM $\{\Pi_j\}$)
  \cite{hol97}.
Since the Holevo information of a single-mode bosonic channel with mean photon
  number constraint is maximized by a coherent state ensemble with a 
  zero-mean circularly-symmetric Gaussian distribution
  \cite{giovannetti04cappureloss}, we have:
\begin{align}
\log_2\kappa+nR&\leq\chi\left((\hat{\rho}^B)^{\otimes n}\right)+1+(\log_2\kappa+nR)\mathbb{P}_e^{(b)}(\mathcal{U})
\end{align}
where $\hat{\rho}^B=\hat{\sigma}^T(\eta\bar{n}_{\mathcal{U}})$, with
  $\hat{\sigma}^T(\bar{n})$ defined in \eqref{eq:TS}.
Now, $\chi(\hat{\rho}^B)=H(\hat{\rho}^B)$ 
  since coherent states are pure, and
  $\chi\left((\hat{\rho}^B)^{\otimes n}\right)=n\left(\log_2(1+\eta\bar{n}_{\mathcal{U}})+\eta\bar{n}_{\mathcal{U}}\log_2\left(1+\frac{1}{\eta\bar{n}_{\mathcal{U}}}\right)\right)$
  is due to the additivity of the Holevo information across the modes of the
  bosonic channels.
This implies:
\begin{align}
\mathbb{P}_e^{(b)}(\mathcal{U})&\geq1-\frac{\log_2(1+\eta\bar{n}_{\mathcal{U}})+\eta\bar{n}_{\mathcal{U}}\log_2\left(1+\frac{1}{\eta\bar{n}_{\mathcal{U}}}\right)+\frac{1}{n}}{\frac{\log_2\kappa}{n}+R}
\end{align}
Since Alice transmits $\omega(\sqrt{n})$ bits in $n$ channel uses,
  her rate is $R=\omega(1/\sqrt{n})$ bits/symbol.
However, $\bar{n}_{\mathcal{U}}=\mathcal{O}(1/\sqrt{n})$, and, as
  $n\rightarrow\infty$, $\mathbb{P}_e^{(b)}(\mathcal{U})$ is bounded away from 
  zero.
Since $\kappa>0$, $\mathbb{P}_e^{(b)}$ is also bounded away from zero 
  when Alice tries to transmit $\omega(\sqrt{n})$ bits in $n$ channel uses
  while beating Willie's heterodyne receiver.
\end{proof}

\ifx\techreport\undefined
\textit{Pure Loss Channel ($N_B=0$) with Quantum-powerful Willie}--- Now
\else
\section{Pure Loss Channel ($N_B=0$) with Quantum-powerful Willie}
Now
\fi
  we prove that Alice and Bob cannot 
  hide their communication from Willie if Willie has a pure loss channel
  from Alice and a choice of a receiver restricted only by the laws of 
  quantum physics.
First, we let Willie pick a receiver that %
  does not necessarily capture
  all the transmitted energy that does not reach Bob's receiver.
Alice uses an arbitrary pure state codebook.
While Willie is oblivious to its structure, we show that Alice must constrain
  her codewords to limit the detection capability of
  Willie's particular receiver.
We then show that this constraint prevents Bob from decoding Alice's 
  transmissions without error, proving the theorem.

\begin{proof}\emph{(Theorem \ref{th:pureloss}).}
Suppose Alice uses a codebook where a state 
  $\Omega_u^A=\bigotimes_{i=1}^n\hat{\rho}_i^A(u)$ encodes message $W_u$ out 
  of $M$ possible messages, with
  $\hat{\rho}^A_i(u)=\ket{\psi_i(u)}\bra{\psi_i(u)}$ and
  $\ket{\psi_i(u)}=\sum_{k=0}^{\nu_i(u)}a^{(i)}_k(u)\ket{k}$ where 
  $\nu_i(u)$ can, in principle, be infinite.
First we analyze Willie's detector and assume that an arbitrary message $W_a$ 
  was transmitted.
Willie captures a fraction of the transmitted 
  energy, $\gamma$, where $0<\gamma\leq 1-\eta$.
Then Willie's hypothesis test reduces to choosing between the states,
\begin{align}
\hat{\rho}_0^{\otimes n}&=\ket{0}\bra{0}^{\otimes n}, \;{\text{and}}\\
\hat{\rho}_1^{\otimes n}&=\bigotimes_{i=1}^n\hat{\rho}^W_i(a)
\end{align}
where $\hat{\rho}^W_i(a)$ is the output state of a pure loss channel with 
  transmissivity $\gamma$ corresponding to an input state $\hat{\rho}^A_i(a)$.
Let Willie use an ideal single photon sensitive direct detection receiver 
  given by positive operator-valued measure (POVM)
  $\left\{\ket{0}\bra{0},\sum_{j=1}^\infty\ket{j}\bra{j}\right\}^{\otimes n}$
  over all $n$ channel uses.
Then Willie's probability of error is
\begin{align}
  \label{eq:pew}\mathbb{P}_e^{(w)}(a)&=\frac{1}{2}\prod_{i=1}^n\bra{0}\hat{\rho}_i^W(a)\ket{0}.
\end{align}
Note that the error is entirely due to the missed codeword detections,
  as Willie's receiver detects vacuum perfectly and never raises a false 
  alarm.

The diagonal elements of $\hat{\rho}_i^W(a)$ expressed in the photon number
  basis are as follows (see Supplement for derivation):
\begin{align}
\label{eq:diag_pl_output}\bra{s}\hat{\rho}_i^W(a)\ket{s}&=\sum_{k=0}^{\nu_i(a)}\left|a_k^{(i)}(a)\right|^2\binom{k}{s}(1-\gamma)^{k-s}\gamma^s
\end{align}
Therefore,
\begin{align}
\bra{0}\hat{\rho}_i^W(a)\ket{0}&=\sum_{k=0}^{\nu_i(a)}\left|a_k^{(i)}(a)\right|^2(1-\gamma)^k\\
&\leq\left|a_0^{(i)}(a)\right| + (1-\left|a_0^{(i)}(a)\right|^2)(1-\gamma)\\
\label{eq:pgf_ub}&=1-\gamma\left(1-\left|a_0^{(i)}(a)\right|^2\right)
\end{align}
Thus, substituting \eqref{eq:pgf_ub} into \eqref{eq:pew} and using the Taylor 
  series expansion of $\log(1-x)$ yields:
\begin{align}
\mathbb{P}_e^{(w)}&\leq\frac{1}{2}\exp\left[-\gamma\sum_{i=1}^n\left(1-\left|a_0^{(i)}\right|^2\right)\right],
\end{align}
implying that Alice must set 
  $\sum_{i=1}^n\left(1-\left|a_0^{(i)}(a)\right|^2\right)=c_a$, 
  with $c_a$ a constant, for every codeword in her codebook with a positive
  probability of being transmitted.
Next we show that the codewords constructed this way are ``too close'' to one 
  another to allow reliable communication.

Let's analyze Bob's receiver.
Denote by $p_u$ the \emph{a priori} probability that $W_u$ is transmitted.
Then, given that $W_u$ is transmitted, the probability of the decoding error
  is the probability of the union of events $\cup_{v=0,v\neq u}^nE_v$, where
  $E_v$ is the event 
  that the received state is decoded as $\hat{W}= W_v$, $v\neq u$.
Let Bob choose a POVM $\{\Lambda_j\}$ that minimizes the average probability
  of error:
\begin{align}
\label{eq:pe_b_1}\mathbb{P}_e^{(b)}&=\inf_{\{\Lambda_j\}}\sum_{u=1}^{M}p_u\mathbb{P}\left(\cup_{v=0,v\neq u}^nE_v|W_u\text{~sent}\right)
\end{align}
Now, any scheme used to transmit a positive number
  of bits has to have at least two messages with positive prior transmission 
  probabilities.
Thus, let's pick a pair of messages $\{W_r,W_s\}$ from Alice's codebook with 
  a positive prior probabilities $\{p_r>0,p_s>0\}$ of transmission. 
Then we have:
\begin{align}
\label{eq:pe_b_lb}\mathbb{P}_e^{(b)}&\geq p_r\mathbb{P}\left(E_s|W_r\text{~sent}\right)+p_s\mathbb{P}\left(E_r|W_s\text{~sent}\right)\\
\label{eq:pe_b_lb1}&=(p_r+p_s)\mathbb{P}_e^{r\leftrightarrow s}
\end{align}
The lower bound in \eqref{eq:pe_b_lb} is due to the exclusion of a non-negative
  elements from the sum in \eqref{eq:pe_b_1}, as well as the events $E_r$ and
  $E_s$ being contained in the unions $\cup_{v=0,v\neq r}^nE_v$ and 
  $\cup_{v=0,v\neq s}^nE_v$, respectively.
In \eqref{eq:pe_b_lb1} we reduced the analytically intractable problem of
  discriminating between many states in \eqref{eq:pe_b_1} to a quantum binary 
  hypothesis test, since
  $\mathbb{P}_e^{r\leftrightarrow s}\equiv\frac{p_r}{p_r+p_s}\mathbb{P}\left(E_s|W_r\text{~sent}\right)+\frac{p_s}{p_r+p_s}\mathbb{P}\left(E_r|W_s\text{~sent}\right)$ 
  is Bob's average probability of error in a scenario where Alice only sends 
  messages $W_r$ and $W_s$ with priors proportional to $p_r$ and $p_s$.
We note that the probabilities are with respect to the POVM $\{\Lambda_j\}$ 
  that minimizes \eqref{eq:pe_b_1} over the entire codebook, and thus may be
  suboptimal for a test between $W_r$ and $W_s$.

Recall that Alice transmits messages by sending codewords through a 
  single mode lossy bosonic channel.
The lower bound on the probability of error in discriminating two received 
  states can be obtained by lower-bounding the probability of error in 
  discriminating two codewords \emph{before} they are sent (this is equivalent 
  to Bob having a channel from Alice with unity transmissivity).
Since the codewords are tensor products of pure states, we can apply the
  Helstrom bound \cite[Eq.~2.34]{helstrom76quantumdetect} 
  for discriminating pure states as follows:
\begin{align}
\label{eq:p_e_bintest}\mathbb{P}_e^{r\leftrightarrow s}&\geq\frac{\left(1-\sqrt{1-\frac{4p_rp_s}{(p_r+p_s)^2}\prod_{i=1}^n\left|\left<\psi_i(r)|\psi_i(s)\right>\right|^2}\right)}{2}
\end{align}
Lower bounding $\prod_{i=1}^n\left|\left<\psi_i(r)|\psi_i(s)\right>\right|^2$
  yields the lower bound on \eqref{eq:p_e_bintest}.
Now, $\prod_{i=1}^n\left|\left<\psi_i(r)|\psi_i(s)\right>\right|^2$ is the 
  fidelity $F(\Omega_r^A,\Omega_s^A)$ between the pure state codewords
  $\Omega_r^A$ and $\Omega_s^A$, which can be represented using the trace 
  distance as follows:
\begin{align}
\label{eq:fid_lb}F(\Omega_r^A,\Omega_s^A)&=1-\frac{1}{4}\|\Omega_r^A-\Omega_s^A\|_1^2\\
\label{eq:triangle_ineq}&\geq1-\frac{(\|\Omega_r^A-\Omega_0\|_1+\|\Omega_s^A-\Omega_0\|_1)^2}{4}
\end{align}
where $\Omega_0=\ket{0}\bra{0}^{\otimes n}$ is the vacuum codeword and 
  \eqref{eq:triangle_ineq} is due to the triangle inequality for trace
  distance.
To lower bound \eqref{eq:triangle_ineq}, we can upper bound the respective trace
  distances between codewords and vacuum using fidelity as follows:
\begin{align}
\|\Omega_r^A-\Omega_0\|_1 &\leq
\sqrt{1-\prod_{i=1}^n\left|\left<0|\psi_i(r)\right>\right|^2}\\
&=\sqrt{1-e^{\sum_{i=1}^n\log(1-(1-\left|\left<0|\psi_i(r)\right>\right|^2))}}\\
\label{eq:exp_c}&\leq\sqrt{1-e^{-(c_r+\mathcal{O}(c_r^2))}}
\end{align}
where \eqref{eq:exp_c} follows from the Taylor series expansion of $\log(1-x)$,
  the fact that
  $\left|\left<0|\psi_i(r)\right>\right|^2)=\left|a_0^{(i)}(r)\right|^2$,
  the fact that Alice has to set 
  $\sum_{i=1}^n\left(1-\left|a_0^{(i)}(r)\right|^2\right)=c_r$ for some constant
  $c_r$ to avoid detection by Willie, and that the square of the sum 
  is greater than the sum of the squares when the sequence contains only
  non-negative numbers.
Analogously, 
\begin{align}
\label{eq:exp_cs}\|\Omega_s^A-\Omega_0\|_1\leq\sqrt{1-e^{-(c_s+\mathcal{O}(c_s^2))}}.
\end{align}
Combining \eqref{eq:pe_b_lb1}, \eqref{eq:p_e_bintest}, \eqref{eq:triangle_ineq},
  \eqref{eq:exp_c} and \eqref{eq:exp_cs} yields:
\begin{widetext}
\begin{align}
\label{eq:pe_b_final_lb}\mathbb{P}_e^{(b)}&\geq\frac{p_r+p_s}{2}\left(1-\sqrt{1-\frac{4p_rp_s}{(p_r+p_s)^2}\left(1-\frac{1}{4}\left(\sqrt{1-e^{-(c_r+\mathcal{O}(c_r^2))}}+\sqrt{1-e^{-(c_s+\mathcal{O}(c_s^2))}}\right)^2\right)}\right)
\end{align}
\end{widetext}
Therefore, by \eqref{eq:pe_b_final_lb}, the probability of error
  is bounded away from zero as the codeword length $n\rightarrow\infty$ and
  reliable covert communication is not possible using pure states
  when Willie has a pure loss channel from Alice and ability to construct an
  ideal single photon sensitive direct detection receiver.
\end{proof}

We have shown above that there exists a quantum measurement that
  Willie can employ to prevent Alice from covertly using a pure loss channel.
However, Alice's situation is not completely hopeless, since the ideal
  direct detection is nearly impossible to realize in practice.

\ifx\techreport\undefined
\textit{Pure Loss Channel ($N_B=0$) with Willie Limited by Practical
  Receiver}---Let 
\else
\section{Pure Loss Channel ($N_B=0$) with Willie Limited by Practical
  Receiver}
Let 
\fi
us reconsider the pure loss channel but assume that Willie's photon
 counting receiver registers a Poisson dark count process with rate $\lambda_d$.
On each symbol interval (channel use) of $\tau$ seconds, the probability of a
dark count at Willie's receiver $p_d \approx \lambda_d \tau$. 
  For instance, $p_d = 10^{-7}$ for a typical superconducting nanowire detector
  with $100$ counts/sec dark count rate and $1$ ns time slots.
The constructive structure of the proof below is similar to that of 
  Theorem \ref{th:thermal}.
  
\begin{proof}\emph{(Theorem \ref{th:dark}).}
Let Alice use a coherent state on-off keying (OOK) modulation $\left\{\pi_i, S_i = |\psi_i\rangle\langle\psi_i |\right\}$, $i = 1, 2$, where $\pi_1 = 1-q$, $\pi_2 = q$, $|\psi_1\rangle = |0\rangle$, $|\psi_2\rangle = |\alpha\rangle$. 
When Alice transmits $|\alpha\rangle$, Bob receives $|\sqrt{\eta}\alpha\rangle$.
Alice and Bob generate a random codebook with each codeword symbol chosen i.i.d. from the above binary OOK constellation.
Since the codebook is kept secret from Willie, Willie observes a sequence of 
  $n$ i.i.d. Bernoulli random variables $\{X_i\}$, $1 \le i \le n$, where $X_i$
  denotes the output of Willie's receiver on the $i^{\text{th}}$ observation.
When Alice is not transmitting (i.e., when $H_0$ is true), the distribution of $X_i$ is $\mathbb{P}_0=\text{Bernoulli}(p_d)$.
When Alice is transmitting a codeword (i.e.~when $H_1$ is true), it is
  $\mathbb{P}_1=\text{Bernoulli}(p_d+q(1-p_d)(1-e^{-(1-\eta)|\alpha|^2}))$ 
  since, as in the proof of Theorem \ref{th:thermal}, Willie 
  captures all of the transmitted energy that does not reach Bob's receiver and
  $\lvert\left<\sqrt{1-\eta}\alpha|0\right>\rvert^2=e^{-(1-\eta)|\alpha|^2}$.

Willie's hypothesis test here is classical and we can thus use the classical 
  relative entropy (CRE) as we do for the AWGN channel in 
  \cite{bash13squarerootjsac,bash12sqrtlawisit} to lower-bound 
  $\mathbb{P}_e^{(w)}$.
CRE is given by  $\mathcal{D}(\mathbb{P}_0\|\mathbb{P}_1)=\sum_{x\in\mathcal{X}}p_0(x)\log\frac{p_0(x)}{p_1(x)}$ where $p_0(x)$ and $p_1(x)$ are the respective densities of 
  $\mathbb{P}_0$ and $\mathbb{P}_1$, and $\mathcal{X}$ is the support of 
  $p_1(x)$.
CRE is additive for independent distributions, and lower-bounds 
  $\mathbb{P}_e^{(w)}\geq\frac{1}{2}-\sqrt{\frac{n}{8}\mathcal{D}(\mathbb{P}_0\|\mathbb{P}_1)}$.
The Taylor series expansion of $\mathcal{D}(\mathbb{P}_0\|\mathbb{P}_1)$ around 
  $|\alpha|^2=0$ yields (via Taylor's Theorem) the following upper bound:
\begin{align}
\label{eq:kl_shotnoise}\mathcal{D}(\mathbb{P}_0\|\mathbb{P}_1)&\leq\frac{(1-p_d)(q(1-\eta)|\alpha|^2)^2}{2p_d}
\end{align}
Thus, to ensure that $\mathbb{P}_e^{(w)}\geq\frac{1}{2}-\epsilon$, Alice can set her
  average symbol power to
\begin{align}
\label{eq:shotnoise_nbar}\bar{n}&=q|\alpha|^2=\frac{4\epsilon}{\sqrt{n}(1-\eta)}\sqrt{\frac{p_d}{1-p_d}}
\end{align}

This allows Alice to transmit $\mathcal{O}(\sqrt{n})$ covert bits reliably 
  to Bob if he also uses a direct detection receiver.
The details of the reliability proof are available in the Supplement.
\end{proof}

Theorems \ref{th:thermal} and \ref{th:dark} suggest that some form of noise 
  in the adversary's measurements, however small, is essential in making LPD 
  communication possible, as LPD communication masquerades as noise.
The nature of the noise appears to be immaterial.
It can come from the thermal 
  environment, be Johnson noise, or be generated locally at the adversary's
  receiver as dark current due to a spontaneous emission process.

Essentially, Alice takes advantage of Willie's measurement noise by 
  transmitting messages, which, when mixed with noise, closely resemble
  the noise that Willie expects to see on his channel when Alice is quiet.
Bob also has to deal with noise in his measurements while decoding,
  but he has a crucial advantage over Willie: his knowledge of the codebook
  allows him to reduce the size of his search space, allowing him to compare
  only the codewords to their received noisy versions.

\ifx\techreport\undefined
\textit{Conclusion}---We 
\else
\section{Conclusion}
We 
\fi
  demonstrated that, provided Willie experiences noise in his measurements
  (either due to thermal noise in the channel or excess local noise in his
  receiver), Alice can transmit $\mathcal{O}(\sqrt{n})$ bits in $n$ channel uses
  to Bob such that Bob's average decoding error probability approaches zero 
  as $n$ gets large while Willie's average probability of detection error is 
  lower-bounded arbitrarily close to $\frac{1}{2}$. 
Surprisingly, this scaling law holds even if Willie obtains a
  quantum-optimal joint-detection measurement over $n$ channel uses and Alice's
  transmissions are subject to thermal noise on the channel.
We also showed that in the absence of any excess noise in Willie's measurements
  (i.e., on a pure loss channel and an ideal detector for Willie), reliable LPD
  communication with coherent state transmission is not possible.

The full converses of Theorems \ref{th:thermal} and \ref{th:dark}
  are open problems that we plan on tackling in the future work.

\bibliography{refs}
\bibliographystyle{apsrev}

\begin{widetext}
\section*{Supplementary material}
\appendix
\subsection{Derivation of \eqref{eq:kl_therm}}
\label{app:der_kl_therm}
Quantum relative entropy
  $D(\rho\|\sigma)\equiv\trace\{\rho(\ln(\rho)-\log(\sigma))\}=-\trace\{\rho(\ln(\sigma))\}-H(\rho)$, 
  where $H(\rho)$ is the von Neumann entropy of the state $\rho$.
Both $\rho_0$ and $\rho_1$ are diagonal in the photon-number basis, 
  which greatly simplifies the calculation of the QRE.
First, let's calculate $-H(\rho_0)$:
\begin{align}
\nonumber-H(\rho_0)&=\trace\left[\left(\sum_{n=0}^\infty \frac{(\eta N_B)^n}{(1+\eta N_B)^{1+n}}\ket{n}\bra{n}\right)\left(\sum_{n=0}^\infty \ln\frac{(\eta N_B)^n}{(1+\eta N_B)^{1+n}}\ket{n}\bra{n}\right)\right]\\
&=\sum_{n=0}^\infty \frac{(\eta N_B)^n}{(1+\eta N_B)^{1+n}}\ln\frac{(\eta N_B)^n}{(1+\eta N_B)^{1+n}}\\
\nonumber&=\frac{1}{1+\eta N_B}\ln\frac{1}{1+\eta N_B}\sum_{n=0}^\infty \left(\frac{\eta N_B}{1+\eta N_B}\right)^n+\\
&\phantom{=}\;+\ln\frac{\eta N_B}{1+\eta N_B}\sum_{n=0}^\infty n\frac{1}{1+\eta N_B}\left(\frac{\eta N_B}{1+\eta N_B}\right)^n\\
\label{eq:th_H_last}&=\ln\frac{1}{1+\eta N_B}+\eta N_B\ln\frac{\eta N_B}{1+\eta N_B}
\end{align}
where \eqref{eq:th_H_last} is due to geometric series 
  $\sum_{n=0}^\infty \left(\frac{\eta N_B}{1+\eta N_B}\right)^n=\left(1-\frac{\eta N_B}{1+\eta N_B}\right)^{-1}$
  and 
  $\sum_{n=0}^\infty n\frac{1}{1+\eta N_B}\left(\frac{\eta N_B}{1+\eta N_B}\right)^n=\eta N_B$
  being the expression for the mean of geometrically distributed random variable
  $X\sim \text{Geom}\left(\frac{1}{1+\eta N_B}\right)$.
We can compute $-\trace[\rho_0\ln(\rho_1)]$ using similar techniques:
\begin{align}
-\trace[\rho_0\ln(\rho_1)]&=-\sum_{n=0}^\infty \frac{(\eta N_B)^n}{(1+\eta N_B)^{1+n}}\ln\frac{((1-\eta)\bar{n}+\eta N_B)^n}{(1+(1-\eta)\bar{n}+\eta N_B)^{1+n}}\\
\nonumber&=-\frac{1}{1+\eta N_B}\ln\frac{1}{1+(1-\eta)\bar{n}+\eta N_B}\sum_{n=0}^\infty \left(\frac{\eta N_B}{1+\eta N_B}\right)^n-\\
&\phantom{=}\;-\ln\frac{(1-\eta)\bar{n}+\eta N_B}{1+(1-\eta)\bar{n}+\eta N_B}\sum_{n=0}^\infty n\frac{1}{1+\eta N_B}\cdot\left(\frac{\eta N_B}{1+\eta N_B}\right)^n\\
&=-\ln\frac{1}{1+(1-\eta)\bar{n}+\eta N_B}-\eta N_B\ln\frac{(1-\eta)\bar{n}+\eta N_B}{1+(1-\eta)\bar{n}+\eta N_B}
\end{align}

\subsection{Derivation of \eqref{eq:exp_het_read_h1} and \eqref{eq:var_het_read_h1}}
To obtain \eqref{eq:exp_het_read_h1} and \eqref{eq:var_het_read_h1} we need 
  the Q representation of the output state observed by Willie 
  $\hat{\rho}^W_i(a)$.
Given the beamsplitter relationship 
  $\hat{w}=\sqrt{\gamma}\hat{a}+\sqrt{1-\gamma}\hat{e}$ between the input 
  modes $a$ and $b$, and the output mode $w$, the Husimi Q function
  $Q_w(\alpha)=\frac{1}{1-\gamma}\int_{\mathbb{C}}Q_a(\beta)Q_b\left(\frac{\alpha-\sqrt{\gamma}\beta}{\sqrt{1-\gamma}}\right)d^2\beta$
  \cite[Eq.~(2.17)]{kim95beamsplitter}.
One of our input modes is the thermal environment $\hat{\rho}^E$ with
  the Q representation
\begin{align}
Q^T(\alpha)=\frac{1}{\pi(1+N_B)}e^{-|\alpha|^2/(1+N_B)}.
\end{align}
The other input mode is Alice's input state 
  $\hat{\rho}^A_i(a)=\ket{\psi_i(a)}\bra{\psi_i(a)}$ with the Q representation
\begin{align}
Q^A_{\ket{\psi_i(a)}}(\alpha)&=\frac{1}{\pi}\sum_{k=0}^{\nu_i(a)}\sum_{l=0}^{\nu_i(a)}b^{(i)}_k(a)\left(b^{(i)}_l(a)\right)^*\frac{(\alpha^*)^k\alpha^l}{\sqrt{k!l!}}e^{-|\alpha|^2},
\end{align}
Using \cite[Eq.~(2.17)]{kim95beamsplitter}, we have:

\begin{align}
Q^W_{\ket{\psi_i(a)}}(\alpha)&=\frac{1}{(1+N_B)(1-\gamma)\pi^2}\int_{\mathbb{C}}e^{-\frac{|\alpha-\sqrt{\gamma}\beta|^2}{(1-\gamma)(1+N_B)}-|\beta|^2}\left(\sum_{k=0}^{\nu_i(a)}\frac{|b^{(i)}_k(a)|^2|\beta|^{2k}}{k!}+\sum_{k=0}^{\nu_i(a)}\sum_{l=0,l\neq k}^{\nu_i(a)}\frac{b^{(i)}_k(a)(b^{(i)}_l(a))^*(\beta^*)^k\beta^l}{\sqrt{k!l!}}\right)d^2\beta\\
\label{eq:q_tonelli}&\begin{aligned}=&\sum_{k=0}^{\nu_i(a)}\frac{|b^{(i)}_k(a)|^2}{(1+N_B)(1-\gamma)\pi^2k!}\int_{0}^\infty\int_0^{2\pi}e^{-\frac{r_{\alpha}^2+\gamma r_{\beta}^2-2\sqrt{\gamma}r_\alpha r_\beta\cos(\theta_\alpha-\theta_\beta)}{(1-\gamma)(1+N_B)}-r_\beta^2}r_\beta^{2k+1}d\theta_\beta dr_\beta\\
&+\sum_{k=0}^{\nu_i(a)}\sum_{l=0,l\neq k}^{\nu_i(a)}\frac{b^{(i)}_k(a)(b^{(i)}_l(a))^*}{(1+N_B)(1-\gamma)\pi^2\sqrt{k!l!}}\int_{0}^\infty\int_0^{2\pi}e^{-\frac{r_{\alpha}^2+\gamma r_{\beta}^2-2\sqrt{\gamma}r_\alpha r_\beta\cos(\theta_\alpha-\theta_\beta)}{(1-\gamma)(1+N_B)}-r_\beta^2}r_\beta^{k+l+1}e^{j(l-k)\theta_\beta}d\theta_\beta dr_\beta
\end{aligned}\\
\label{eq:q_bessel}&\begin{aligned}=&\sum_{k=0}^{\nu_i(a)}\frac{2|b^{(i)}_k(a)|^2}{(1+N_B)(1-\gamma)\pi k!}\int_{0}^\infty e^{-\frac{r_{\alpha}^2+(1+(1-\gamma) N_B)r_\beta^2}{(1-\gamma)(1+N_B)}}I_0\left(\frac{2\sqrt{\gamma}r_\alpha r_\beta}{(1-\gamma)(1+N_B)}\right)r_\beta^{2k+1} dr_\beta\\
&+\sum_{k=0}^{\nu_i(a)}\sum_{l=0,l\neq k}^{\nu_i(a)}\frac{2b^{(i)}_k(a)(b^{(i)}_l(a))^*}{(1+N_B)(1-\gamma)\pi\sqrt{k!l!}}\int_{0}^\infty e^{-\frac{r_{\alpha}^2+(1+(1-\gamma) N_B)r_\beta^2}{(1-\gamma)(1+N_B)}}I_{l-k}\left(\frac{2\sqrt{\gamma}r_\alpha r_\beta}{(1-\gamma)(1+N_B)}\right)r_\beta^{k+l+1}e^{j(l-k)\theta_\alpha} dr_\beta
\end{aligned}%
\end{align}
where in \eqref{eq:q_tonelli} we substituted the polar form of complex variables
  $\alpha=r_\alpha e^{j\theta_\alpha}$ and $\beta=r_\beta e^{j\theta_\beta}$ 
  as well as changed the order of integration and summation.
The latter is justified by Tonelli's theorem, as Q-functions are positive.
\eqref{eq:q_bessel} is due to the integral-based definition of the modified 
  Bessel function of the first kind 
  $I_n(z)=\frac{1}{\pi}\int_0^\pi e^{z\cos \theta}\cos(n\theta)d\theta$.

We obtain the expected squared magnitude of heterodyne detector reading when 
  Alice transmits $\hat{\rho}_i^A(a)$ using \eqref{eq:q_bessel}:
\begin{align}
\mathbb{E}[|y_i|^2]&=\int_{\mathbb{C}}|\alpha|^2Q^W_{\ket{\psi_i(a)}}(\alpha)d^2\alpha\\
\label{eq:evalQ}&=\int_{0}^\infty\int_{0}^{2\pi}r_\alpha^3Q^W_{\ket{\psi_i(a)}}(r_\alpha e^{j\theta_\alpha})d\theta_\alpha dr_\alpha\\
\label{eq:e_yi}&=\sum_{k=0}^{\nu_i(a)}|b^{(i)}_k(a)|^2(1+(1-\gamma)N_B+\gamma k)\\
&=1+(1-\gamma)N_B+\gamma\bar{n}_i(a).
\end{align}
When evaluating \eqref{eq:evalQ} we note that the second (double)
  summation in \eqref{eq:q_bessel} is zero because 
  $\int_0^{2\pi}e^{j(l-k)\theta_\alpha}d\theta_\alpha=0$ when $l\neq k$.
Thus, we only need to integrate the first summation in \eqref{eq:q_bessel}.
We substitute the 
  summation-based definition of the modified Bessel function of the first
  kind $I_0(z)=\sum_{m=0}^\infty \frac{(z/2)^{2m}}{(m!)^2}$ and
  change in the order of summation and integration, using Tonelli's theorem
  to justify the latter step since the arguments in summations are non-negative.
The the integrals with respect to $r_{\alpha}$ and $r_\beta$ take a form with
  the following solution \cite[Eq.~(3.326.2)]{gr07tables}: 
  $\int_0^\infty x^me^{-cx^n}dx=\frac{\Gamma(\kappa)}{nc^\kappa}$ where 
  $\kappa=(m+1)/n$.
Finally, to arrive at \eqref{eq:e_yi} we use the identity 
  $\sum_{m=0}^\infty \frac{r^m(m+n)!}{m!}=n!\sum_{m=0}^\infty r^m \binom{m+n}{m}=\frac{n!}{(1-r)^{n+1}}$
  which is valid for any $r$ satisfying $0\leq r<1$ as is our case.
  
Similarly, the second moment of the square magnitude of heterodyne detector 
  reading when Alice transmits $\hat{\rho}_i^A(a)$ is obtained as follows:
\begin{align}
\mathbb{E}[|y_i|^4]&=\int_{\mathbb{C}}|\alpha|^4Q^W_{\ket{\psi_i(a)}}(\alpha)d^2\alpha\\
&=\int_{0}^\infty\int_{0}^{2\pi}r_\alpha^5Q^W_{\ket{\psi_i(a)}}(r_\alpha e^{j\theta_\alpha})d\theta_\alpha dr_\alpha\\
&=\sum_{k=0}^{\nu_i(a)}|b^{(i)}_k(a)|^2(\gamma^2k^2+2(1+(1-\gamma)N_B)^2+4\gamma(1-\gamma)(1+N_B)k)\\
&=\gamma^2\mu^{(2)}_i(a)+2(1+(1-\gamma)N_B)^2+4\gamma(1-\gamma)(1+N_B)\bar{n}_i(a)
\end{align}
where $\mu^{(2)}_i(a)=\sum_{k=0}^{\nu_i(a)}k^2|b^{(i)}_k(a)|^2$.
The variance of the squared magnitude of heterodyne detector reading when Alice
  transmits $\hat{\rho}_i^A(a)$ is then:
\begin{align}
\Var[|y_i|^2]&=\gamma^2\sigma^2_i(a)+c_1\bar{n}_i(a)+c_2
\end{align}
where $\sigma^2_i(a)=\mu^{(2)}_i(a)-(\bar{n}_i(a))^2$ denotes the 
  photon number variance of $\hat{\rho}_i^A(a)$, and
  $c_1=2\gamma((2+N_B)(1-\gamma)-1)$, $c_2=(1+(1-\gamma)N_B)^2$.

\subsection{Derivation of \eqref{eq:diag_pl_output}}
A beamsplitter can be described as a unitary transformation $U_{BS}$ from two
  input modes to two output modes.
In our scenario, the inputs are Alice's input state 
  $\hat{\rho}^A=\ket{\psi}^{A}\prescript{A}{}{\bra{\psi}}$ and vacuum  
  environment $\hat{\rho}^E=\ket{0}^{E}\prescript{E}{}{\bra{0}}$.
The outputs are Willie's output state $\hat{\rho}^W$ and
  Bob's output state $\hat{\rho}^B$.
First, suppose Alice transmits a number state $\ket{\psi}^A=\ket{k}^A$.
Then the inputs and outputs of a beamsplitter with transmissivity $\gamma$ are 
  related as follows:
\begin{align}
U_{BS}\ket{k}^A\ket{0}^E&=\sum_{m=0}^k\sqrt{\binom{k}{m}\gamma^m(1-\gamma)^{k-m}}\ket{m}^W\ket{k-m}^B.
\end{align}
Now suppose that Alice transmits an arbitrary pure state expressed in the
  number basis as follows: $\ket{\psi}^A=\sum_{k=0}^\infty a_k\ket{k}^A$.
Since $U_{BS}$ is a linear transformation,
\begin{align}
U_{BS}\left(\sum_{k=0}^\infty a_k\ket{k}^A\right)\ket{0}^E&=\sum_{k=0}^\infty a_k\sum_{m=0}^k\sqrt{\binom{k}{m}\gamma^m(1-\gamma)^{k-m}}\ket{m}^W\ket{k-m}^B\equiv\ket{\psi}^{WB}
\end{align}
with the output state 
  $\hat{\rho}^{WB}=\ket{\psi}^{WB}\prescript{WB}{}{\bra{\psi}}$.
However, we desire only Willie's output state $\hat{\rho}^W$, which we obtain
  using the partial trace over the Bob's output state:
\begin{align}
\hat{\rho}^W&=\trace_B\left[\ket{\psi}^{WB}\prescript{WB}{}{\bra{\psi}}\right]\\
&=\sum_{n=0}^\infty\prescript{B}{}{\langle n|\psi\rangle}^{WB}\prescript{WB}{}{\langle \psi|n\rangle}^{B}
\end{align}
where
\begin{align}
\prescript{B}{}{\langle n|\psi\rangle}^{WB}&=\sum_{k=0}^\infty a_k\sum_{m=0}^k\sqrt{\binom{k}{m}\gamma^m(1-\gamma)^{k-m}}\ket{m}^W\prescript{B}{}{\langle n|k-m\rangle}^B\\
\label{eq:orthnum}&=\sum_{k=0}^\infty a_k\sqrt{\binom{k}{n}\gamma^{k-n}(1-\gamma)^{n}}\ket{k-n}^W
\end{align}
with \eqref{eq:orthnum} due to the orthonormality of number states.
Thus,
\begin{align}
\bra{s}\hat{\rho}^W\ket{s}&=\sum_{n=0}^\infty |a_n|^2\binom{n}{s}\gamma^s(1-\gamma)^{n-s}
\end{align}
where we use the convention that $\binom{a}{b}=0$ when $a<b$.

\subsection{Reliability of LPD Communication Using OOK Modulation}
\label{app:shot_noise_rel}
\begin{figure}[h]
\vspace{0.10in}
\begin{center}
\begin{picture}(45,33)
\put(0,30.5){\makebox(5,5){$\ket{0}$}}
\put(0,0.5){\makebox(5,5){$\ket{\alpha}$}}
\put(45,30.5){\makebox(5,5){$0$}}
\put(45,0.5){\makebox(5,5){$1$}}
\put(5,3){\vector(4,3){40}}
\put(5,33){\vector(4,-3){40}}
\put(5,3){\vector(1,0){40}}
\put(5,33){\vector(1,0){40}}
\put(30,19){\makebox(5,5){$\delta$}}
\put(22,33){\makebox(5,5){$1-\beta$}}
\put(30,12){\makebox(5,5){$\beta$}}
\put(22,-2){\makebox(5,5){$1-\delta$}}
\end{picture}
\end{center}
\vspace{-0.10in}
\caption{The binary asymmetric channel between Alice and Bob.  Input probabilities are $p(\ket{0})=1-q$ and $p(\ket{\alpha})=q$.  Transition probabilities are $\delta=e^{-\eta |\alpha|^2}(1-p_b)$ and $\beta=p_b$.}
\label{fig:bac}
\vspace{-0.10in}
\end{figure}
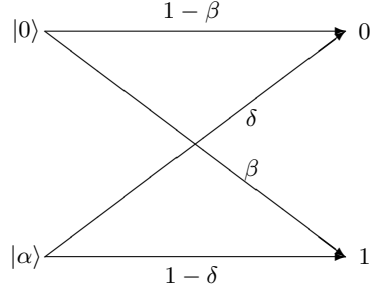

Dark current in Bob's receiver induces a binary asymmetric channel (BAC) between
  Alice and Bob depicted in Figure \ref{fig:bac}.
Since the channel between Alice and Bob is a classical \emph{discrete 
  memoryless channel} (DMC), by \cite[Th.~5.6.1]{gallager68IT} and the
  discussion that follows it in \cite{gallager68IT},
  Bob's average probability of decoding
  error $\mathbb{P}_e^{(b)}$ can be upper-bounded as follows:
\begin{align}
\label{eq:p_e2}\mathbb{P}_e^{(b)}&\leq e^{-n(E_{0}(s)-sR)},
\end{align}
where $n$ is the size of the codeword, $p_b$ is Bob's receiver dark click
  probability, $R$ is the coding rate, $0\leq s\leq 1$,
  and $E_{0}(s)$ is defined as follows:
\begin{align}
E_0(s)&=-\ln\left[(1-p_b)\left(1-q\left(1-e^{-\frac{\eta|\alpha|^2}{1+s}}\right)\right)^{1+s}+\left((1-q)p_b^{1/(1+s)}+q\left(1-(1-p_b)e^{-\eta|\alpha|^2}\right)^{1/(1+s)}\right)^{1+s}\right]
\end{align}
However, the Taylor series expansion around $|\alpha|^2=0$ has a zero 
  first-order term:
\begin{align}
E_0(s)&=\frac{(1-q)q(1-p_b)s\eta^2|\alpha|^4}{2p_b(1+s)}+\mathcal{O}(|\alpha|^6)
\end{align}
Therefore, Alice and Bob have to set their per-symbol mean photon number
  $|\alpha|^2=\omega(1/\sqrt{n})$ to upper-bound Bob's probability
  of decoding error by an arbitrary $\delta>0$.
However, recall that to prevent the detection by Willie, they must set the 
  mean photon number $\bar{n}=q|\alpha|^2$ to \eqref{eq:shotnoise_nbar}.
Thus, using a method similar to the one described in
  \cite[App.~A]{bash13squarerootjsac}, they can construct a covert codebook in 
  two stages.
First, Alice and Bob randomly select the symbol periods that they will use
  for their transmission by flipping a biased coin $n$ times and selecting
  the $i^{\text{th}}$ symbol period with probability $c/\sqrt{n}<1$
  for some constant $c$.
Denote the number of selected symbol periods by $\tau$ and note that
  mean $\bar{\tau}=c\sqrt{n}$.
Second, set
\begin{align}
|\alpha|^2&=\frac{4\epsilon n}{\tau\sqrt{n}(1-\eta)}\sqrt{\frac{p_d}{1-p_d}}
\end{align}
and generate the codebook with codewords of length $\tau$ on the selected
  $\tau$ symbol periods.
Since the symbol location selection is independent of both the symbol and the
  channel noise, the analysis leading to \eqref{eq:kl_shotnoise} applies.
Covert communication criterion \eqref{eq:shotnoise_nbar} is 
  satisfied, and $|\alpha|^2=\omega(1/\sqrt{n})$ with high probability,
  ensuring reliable transmission of $\mathcal{O}(\sqrt{n})$ covert bits
  from Alice to Bob.

\end{widetext}
\end{document}